# Charge-induced phase separation in lipid membranes

Hiroki Himeno[a†], Naofumi Shimokawa[a†], Shigeyuki Komura[b], David Andelman[c], Tsutomu Hamada[a*], and Masahiro Takagi[a]

**Abstract**
The phase separation in lipid bilayers that include negatively charged lipids is examined experimentally. We observed phase-separated structures and determined the membrane miscibility temperatures in several binary and ternary lipid mixtures of unsaturated neutral lipid, dioleoylphosphatidylcholine (DOPC), saturated neutral lipid, dipalmitoylphosphatidylcholine (DPPC), unsaturated charged lipid, dioleoylphosphatidylglycerol (DOPG$^{(-)}$), saturated charged lipid, dipalmitoylphosphatidylglycerol (DPPG$^{(-)}$), and cholesterol. In binary mixtures of saturated and unsaturated charged lipids, the combination of the charged head with the saturation of hydrocarbon tail is a dominant factor for the stability of membrane phase separation. DPPG$^{(-)}$ enhances phase separation, while DOPG$^{(-)}$ suppresses it. Furthermore, the addition of DPPG$^{(-)}$ to a binary mixture of DPPC/cholesterol induces phase separation between DPPG$^{(-)}$-rich and cholesterol-rich phases. This indicates that cholesterol localization depends strongly on the electric charge on the hydrophilic head group rather than on the ordering of the hydrocarbon tails. Finally, when DPPG$^{(-)}$ was added to a neutral ternary system of DOPC/DPPC/Cholesterol (a conventional model of membrane rafts), a three-phase coexistence was produced. We conclude by discussing some qualitative features of the phase behaviour in charged membranes using a free energy approach.

**Introduction**
One of the major components of cell membranes is their lipid bilayer composed of a mixture of several phospholipids, all having a hydrophilic head group and two hydrophobic tails. Recently, a number of studies have investigated heterogeneities in lipid membranes in relation with the lipid raft hypothesis[1,2]. Lipid rafts are believed to function as a platform on which proteins are attached during signal transduction and membrane trafficking[3]. It is commonly believed (but still debatable) that the raft domains are associated with phase separation that takes place in multi-component lipid membranes[4].

In order to reveal the mechanism of phase separation in lipid membranes, giant unilamellar vesicles (GUV) consisting of mixtures of lipids and cholesterol have been used as model biomembranes[5-7]. In particular, studies of phase separation and membrane dynamics have been performed on such GUV consisting of saturated lipids, unsaturated lipids and cholesterol[8]. Multi-component membranes phase separate into domains rich in saturated lipids and cholesterol, whereas the surrounding fluid phase is composed largely of unsaturated lipids. The essential origin of this lateral phase separation was argued to be the hydrophobic interactions between acyl chains of lipid molecules.

In the past, most of the studies have investigated the phase separation in uncharged model membranes[9-11]. However, biomembranes also include charged lipids, and, in particular, phosphatidylglycerol (PG$^{(-)}$) is found with high fractions in prokaryotic membranes. In this respect it is worth mentioning that in *Staphylococcus aureus* the PG$^{(-)}$ membranal fraction is as high as 80%, whereas the *Escherichia coli* membrane includes 15% of PG$^{(-)}$[12]. Although the charged lipid fraction in eukaryotic plasma membranes is lower, its sub-cellular organelles such as mitochondria and lysosome are enriched with several types of charged lipids[13]. For example, mitochondria inner membrane includes 20% of charged lipids such as cardiolipin (CL$^{(-)}$), phosphatidylserine (PS$^{(-)}$) and PG$^{(-)}$[14,15]. It is indispensable to include the effect of electrostatic interactions on the phase behavior in biomembranes. To emphasize even further the key role played by the charges, we note that membranes composed of a binary mixture of charged lipids was reported to undergo a phase separation induced by addition of salt, even when the two lipids have same hydrocarbon tail[16-18]. For this charged lipid mixture, the segregation is mediated only by the electrostatic interaction between the lipids and the electrolyte.

In related studies, Shimokawa *et al*[19,20] studied mixtures consisting of neutral saturated lipid (DPPC), negatively charged unsaturated lipid (DOPS$^{(-)}$) and cholesterol. The main result is the suppression of the phase separation due to electrostatic interactions between the charged DOPS$^{(-)}$ lipids. Two other relevant studies are worth mentioning. Vequi-Suplicy *et al*[21], reported the suppression of phase separation using other charged unsaturated lipids, and more recently Blosser *et al*[22] investigated the phase diagram and miscibility temperature in mixtures containing charged lipids. However, the effect of electric charge on the phase behaviour in lipid/cholesterol mixtures have not been addressed so far systematically.

In the present study, we investigate the physicochemical properties of *model* membranes containing various mixtures of charged lipids, with the hope that the study will enhance our understanding of biomembranes *in-vivo*, which are much more complex. We examine the electric charge effect on the phase behaviour using fluorescent microscopy and confocal laser scanning microscopy. In addition, the salt screening effect on charged membranes is explored. We discuss these effects in three stages starting from the simpler one. First, the phase diagram in charged *binary* mixtures of unsaturated and saturated lipids is presented. Second, we investigate the phase behaviour in *ternary* mixtures consisting of saturated lipids (charged and neutral) and cholesterol. And third, we include the change of phase behaviour when a charged saturated lipid is added as a *fourth* component to a ternary mixture of neutral saturated and unsaturated lipids and cholesterol. We conclude by discussing qualitatively the phase behaviour of charged membranes using a free energy modeling. The counterion concentration adjacent to the charged membrane is calculated in order to explore the relation between the electric charge and the ordering of hydrocarbon tail.

## Materials and methods
### Materials

Neutral unsaturated lipid dioleoyl-*sn*-glycero-3-phosphocholine (DOPC, with chain melting temperature, $T_m$= -20℃), neutral saturated lipid dipalmitoyl-*sn*-glycero-3-phosphocholine (DPPC, $T_m$ = 41℃), negatively charged unsaturated lipid 1,2-dioleoyl-*sn*-glycero-3-phospho-(1'-rac-glycerol) (sodium salt) (DOPG$^{(-)}$, $T_m$=-18℃), negatively charged saturated lipid 1,2-dipalmitoyl-*sn*-glycero-3-phospho-(1'-rac-glycerol) (sodium salt) (DPPG$^{(-)}$, $T_m$= 41℃), and cholesterol were obtained from Avanti Polar Lipids (Alabaster, AL). BODIPY labelled cholesterol (BODIPY-Chol) and Rhodamine B 1,2-dihexadecanoyl-*sn*-glycero-3-phosphoethanolamine (Rhodamine-DHPE) were purchased from Invitrogen (Carlsbad, CA). Deionized water was obtained from a Millipore Milli-Q purification system. We chose phosphatidylcholine (PC) as the neutral lipid head and phosphatidylglycerol (PG) as the negatively charged lipid head because the chain melting temperature of PC and PG lipids having the same acyl tails, is almost identical. In cellular membranes, PC is the most common lipid component, and PG is highly representative among charged lipids.

### Preparation of giant unilamellar vesicles

Giant unilamellar vesicles (GUVs) were prepared by gentle hydration method. Lipids and fluorescent dyes were dissolved in 2:1 (vol/vol) chloroform/methanol solution. The organic solvent was evaporated under a flow of nitrogen gas, and the lipids were further dried under vacuum for 3h. The films were hydrated with 5 μL deionized water at 55 ℃ for 5 min (pre-hydration), and then with 200 μL deionized water or NaCl solution for 1-2 h at 37℃. The final lipid concentration was 0.2 mM. Rhodamine-DHPE and BODIPY-Chol concentrations were 0.1 μM and 0.2 μM, respectively.

## Microscopic observation

The GUV solution was placed on a glass coverslip, which was covered with another smaller coverslip at a spacing of ca. 0.1 mm. We observed the membrane structures with a fluorescent microscope (IX71, Olympus, Japan) and a confocal laser scanning microscope (FV-1000, Olympus, Japan). In the present study, Rhodamine-DHPE and BODIPY-Chol were used as fluorescent dyes. Rhodamine-DHPE labels the lipid liquid phase, whereas BODIPY-Chol labels the cholesterol-rich one. A standard filter set U-MWIG with excitation wavelength, $\lambda_{ex}$=530–550nm, and emission wavelength, $\lambda_{em}$=575 nm, was used to monitor the fluorescence of Rhodamine-DHPE, and another filter, U-MNIBA with $\lambda_{ex}$=470–495 nm and $\lambda_{em}$=510-550 nm, was used for the BODIPY-Chol dye. The sample temperature was controlled with a microscope stage (type 10021, Japan Hitec).

## Measurement of miscibility temperature

The miscibility temperature corresponds to the boundary between one- and two-phase regions. It is defined as the phase separation point at which more than 50% of the phase-separated domains have disappeared upon heating. The temperature was increased from room temperature to the desired temperature by 10 °C/min, and a further delay of 5 min was used in order to approach the equilibrium state. We then measured the percentage of vesicles that were in the two-phase coexisting region. If the percentage of such two-phase vesicles was over 50%, the temperature was further increased by 2 °C. We continued this procedure until the percentage of two-phase vesicles decreased below 50%.

## Results

### Binary lipid mixtures

First, we focus on the effect of charges on the phase separation of binary unsaturated/saturated lipid mixtures. We use neutral unsaturated lipid DOPC, neutral saturated lipid DPPC, negatively unsaturated lipid DOPG$^{(-)}$, and negatively saturated lipid DPPG$^{(-)}$ (see Table 1). We observed the phase separation and measured the miscibility temperatures in three different binary mixtures: DOPC/DPPC, DOPC/DPPG$^{(-)}$, and DOPG$^{(-)}$/DPPC. Figure 1(A) shows the phase behaviour in these three binary mixtures taken for three temperatures: $T$ = 22°C, 30°C and 40°C. Each of the images was taken by superimposing several pictures at a slightly different focus position of the confocal laser scanning microscope. At room temperature (22°C), all three mixtures exhibit a phase separation (images 7, 8, and 9). The red regions indicate liquid-disordered phase ($L_d$) that includes large amount of the unsaturated lipid, while the dark regions represent solid-ordered phase ($S_o$) that is rich in the saturated lipid. When the temperature was raised to 30°C, the phase separation of DOPG$^{(-)}$/DPPC disappears (image 6). On the other hand, the two other mixtures (DOPC/DPPC and DOPC/DPPG$^{(-)}$) still kept the phase-separated structure (images 4 and 5). As the temperature was further increased to 40°C, the DOPC/DPPC mixture also becomes homogeneous (image 1), while the DOPC/DPPG$^{(-)}$ mixture still retains its phase-separated structure at the same temperature (image 2). Thus, DOPC/DPPG$^{(-)}$ mixture shows the highest miscibility temperature of all studied systems. Note that a similar phase-separated structure was reported in binary mixtures of egg sphingomyelin (eSM)/DOPG$^{(-)[21,23]}$.

Miscibility temperatures of binary mixtures are summarized in Fig. 1(B). The filled circles denote the neutral lipid mixture, DOPC/DPPC. We also examined charged binary mixtures of two negatively charged lipids, DOPG$^{(-)}$/DPPG$^{(-)}$. Miscibility temperatures (data not shown) were quite similar to those of neutral DOPC/DPPC mixtures. This implies that the phase separation behavior is determined by the interaction between hydrocarbon tails in mixtures consisting of the same lipid head group. When the neutral unsaturated lipid (DOPC) was replaced with charged unsaturated lipid (DOPG$^{(-)}$), the miscibility temperature in the DOPG$^{(-)}$/DPPC mixture (denoted by filled triangles) becomes lower as compared with a neutral lipid mixture, DOPC/ DPPC. In other words, the phase separation is suppressed when a negatively charged unsaturated lipid is included. This result is consistent with previous studies performed on lipid mixtures containing negatively charged unsaturated lipids[19,21,22,23]. At higher concentrations of DPPC, phase-separated domains could not be observed for mixtures of DOPG$^{(-)}$/DPPC=20:80 and 10:90, because stable vesicle formation was prevented by the larger amount of DPPC.

We also replaced the neutral saturated lipid, DPPC, with negatively charged saturated lipid, DPPG$^{(-)}$. In the DOPC/DPPG$^{(-)}$ mixture, the miscibility temperature (denoted by filled squares in Fig. 1(B)) increases significantly as compared with the neutral system. In particular, we can see that a maximum in the miscibility temperature appears in the phase diagram around 50% relative concentration of the saturated lipid. Interestingly, at DOPC/DPPG$^{(-)}$=50:50, the miscibility temperature of about 44℃ was higher than 41℃ of the DPPG$^{(-)}$ chain melting temperature (Table 1). Thus, the phase separation is enhanced in mixtures containing negatively charged saturated lipid (DPPG$^{(-)}$). This result should be contrasted with the phase behaviour of the DOPG$^{(-)}$/DPPC charged/neutral mixture. We will further elaborate on such a phase behaviour in the discussion section.

The phase behaviour of charged membranes is also investigated in presence of salt (10mM NaCl solution) for various charged/neutral mixtures. The miscibility temperatures for DOPG$^{(-)}$/DPPC and DOPC/DPPG$^{(-)}$ with NaCl solutions are indicated by open triangles and squares, respectively, in Fig. 1(B). The phase separation was enhanced by the addition of salt for DOPG$^{(-)}$/DPPC, which is in agreement with the previous findings[19,21]. On the other hand, the phase separation of DOPC/DPPG$^{(-)}$ with NaCl was suppressed. It seems that the phase behaviour in charged membranes with salt approaches that of the neutral mixture, DOPC/DPPC. This is consistent with the fact that salt screens the electrostatic interactions of the charged DOPG$^{(-)}$ and DPPG$^{(-)}$ lipids.

**Ternary lipid/cholesterol mixtures**

In general, cholesterol prefers to be localized in the saturated lipid-rich phase rather than in the unsaturated lipid-rich one. However, the localization of cholesterol also depends strongly on the structure of lipid head group[24]. We investigated the localization of cholesterol and the resulting phase behaviour in ternary mixtures composed of neutral saturated lipid, negatively charged saturated lipid and cholesterol, such as DPPC/DPPG$^{(-)}$/Chol. The effect of the hydrocarbon tail was excluded by using lipids with the same acyl chain.

The phase behavior of DPPC/DPPG$^{(-)}$/Chol mixtures for Milli Q water and NaCl aqueous solutions is summarized in Fig. 2. Although the cholesterol solubility limit in phospholipid membranes is about 60%, we show the results for Chol>60% to emphasize the phase boundary, especially in the case of Milli Q water. For membranes consisting only of neutral lipids (DPPC/Chol=80:20), the phase separation was not observed at room temperature, as shown in image 1 of Fig. 2. In DPPC/Chol binary mixture, however, it was reported that the nanoscopic domains are formed even though they cannot be detected by optical microscopes[25]. On the other hand, when we replaced a fraction of the DPPC with negatively charged lipid DPPG$^{(-)}$, DPPC/DPPG$^{(-)}$/Chol = 40:40:20, a stripe-shaped domain was observed using Rhodamine-DHPE fluorescent dye as shown in image 2 of Fig. 2. Since the stripe-shaped domain has an anisotropic shape, this is a strong indication that the domain is in the $S_o$ phase. The phase behavior of DPPC/DPPG$^{(-)}$/Chol mixtures in Milli Q water is summarized in the left diagram of Fig. 2. For higher concentrations of DPPC or cholesterol, two-phase vesicles were not observed or rarely observed (open circle). On the other hand, their percentage clearly increases with the DPPG$^{(-)}$ concentration (filled circle).

Three experimental findings led us to conclude that red and dark regions in the fluorescence images represent, respectively, DPPC/Chol-rich and DPPG$^{(-)}$-rich phases. (i) The domain area (dark region) became larger as the percentage of DPPG$^{(-)}$ was increased, as shown in Fig. 3(C). (ii) While the homogeneous phase is stable for DPPC/Chol mixtures, DPPG$^{(-)}$/Chol mixtures show a phase separation. Therefore, cholesterol molecules mix easily with DPPC but not with

DPPG$^{(-)}$. (iii) We used BODIPY-Chol as a fluorescent probe that usually favors the cholesterol-rich phase. The BODIPY-Chol was localized in the red regions stained by Rhodamine-DHPE (the data is not shown). Although the bulky BODIPY-Chol may not behave completely like cholesterol, BODIPY-Chol is partitioned into Chol-rich phase in all our experiments[26]. In addition, we also observed the phase behaviors without BODIPY-Chol, and the observed results did not change in any significant way. Thus, we think that bulky BODIPY-Chol plays a rather minor role in our study.

Since most of the cholesterol is included in the DPPC/Chol-rich region, the DPPC/Chol-rich region is identified as a liquid-ordered ($L_o$) phase. In contrast, the DPPG$^{(-)}$-rich domain is in an $S_o$ phase, because its domain shape is not circular but rather stripe-like. We also note that without cholesterol, a membrane composed of pure DPPG$^{(-)}$ will be in an $S_o$ phase at room temperature (lower than its chain melting temperature, $T_m$=41 °C). Our results indicate that DPPG$^{(-)}$ tends to repel DPPC and cholesterol. In other words, the interaction between the head groups of the lipids affects the localization of cholesterol. Furthermore, as the fraction of DPPG$^{(-)}$ of DPPC/ DPPG$^{(-)}$/cholesterol membranes increases, the corresponding miscibility temperature also increases continuously (Fig. 3A). For systems with DPPG$^{(-)}$ percentage of over 30%, two-phase coexistence was observed even above the chain melting temperature of DPPG$^{(-)}$ (Table 1). It implies that the head group interaction of DPPG$^{(-)}$ makes a large contribution to stabilize the phase structure. We will further discuss this point in the discussion section.

We now turn to the addition of salt and its effect on the phase behaviour. The phase-separated regions with 1mM and 10mM of NaCl are indicated in Fig. 2. As the salt concentration is increased, the phase separation tends to be suppressed. This can be understood because DPPG$^{(-)}$ is screened in presence of salt and approaches the behaviour of the neutral DPPC. This observation is qualitatively consistent with the result for DOPC/ DPPG$^{(-)}$ mixtures shown in Fig. 1. For fixed amount of Chol=20%, we measured the percentage of two-phase vesicles and the area percentage of the $S_o$ phase. The results are summarized in Fig. 3(B) and (C). From Fig. 3(B) we can see that the addition of salt decreases the percentage of domain formation. Also, the phase separation is enhanced in the region where a large amount of DPPG$^{(-)}$ is included, as DPPG$^{(-)}$ molecules tend to exclude the cholesterol.

A further finding is shown in Fig. 3(C), where it can be seen that the area fraction of $S_o$ phase decreases by the addition of the salt. Since salt screens the DPPG$^{(-)}$ charge, DPPG$^{(-)}$ tends to be incorporated into the $L_o$ phase, similarly to what is seen for neutral DPPC.

## Four-component mixtures of lipid and cholesterol

From the results of ternary mixtures, we conclude that cholesterol prefers to be localized in the neutral DPPC-rich domains rather than in the DPPG$^{(-)}$-rich ones.

Next, we investigated four-component mixtures of DOPC/DPPC/DPPG$^{(-)}$/Chol. Previously, a number of studies have used the mixtures of DOPC/DPPC/Chol as a biomimetic system related to modelling of rafts[8]. In these mixtures, unsaturated lipids (DOPC) form an $L_d$-phase, whereas domains rich in saturated lipids (DPPC) and cholesterol form an $L_o$-phase. Aiming to reveal the effect of charge on the $L_d$/$L_o$ phase separation, we replace a fraction of the DPPC component in the DOPC/DPPC/Chol mixture with negatively charged saturated lipid, DPPG$^{(-)}$. We also screen head group charge by adding salt, and examined how the charged lipid, 4$^{th}$ component, affects phase organization of the ternary mixture.

For ternary mixtures with DOPC/DPPC/Chol = 40:40:20 (without the charged lipid), a phase separation is observed, Fig. 4(A1), using the Rhodamine-DHPE dye (red color) and the BODIPY-Chol dye (green color). The circular green domains are rich in DPPC and cholesterol, inferring an $L_o$ phase, while the red region is a DOPC-rich ($L_d$) phase. When half of DPPC was replaced by the charged DPPG$^{(-)}$, a distinct phase separation (three-phase coexistence) was observed in the four-component mixture, DOPC/DPPC/DPPG$^{(-)}$/Chol = 40:20:20:20, as shown in Fig. 4(A2). The black regions that appear inside the green domains, contain a large amount of DPPG$^{(-)}$ as is the case of ternary mixtures. Because this black region excludes any fluorescent dyes, the DPPG$^{(-)}$-rich region is inferred as the S$_o$ phase. We consider that the observed three-phase coexistence is equilibrated, since the three-phase coexistence reappears at the same temperature when the system is heated and cooled again.

Moreover, for ternary mixtures of DOPC/DPPG$^{(-)}$/Chol = 40:40:20 without DPPC, a coexistence between S$_o$ and L$_d$ phases is observed as shown in Fig. 4(A3). The phase diagram of DOPC/DPPC/DPPG$^{(-)}$ for fixed Chol = 20% presented in Fig. 4(B) shows that the phase-separation strongly depends on the DPPG$^{(-)}$ concentration. The boundary between the L$_o$/S$_o$ and L$_d$/S$_o$ coexistence is not marked on the phase diagram, because from optical microscopy it was not possible to distinguish between the L$_o$ and L$_d$ phases. But the region where S$_o$ coexists with either L$_o$ or L$_d$ is indicated as light grey region in the phase diagram.

Interestingly, at DOPC/DPPC/DPPG$^{(-)}$/Chol = 40:15:25:20, a transition between two-phase and three-phase coexistence was driven by adding salt, as is shown in the images of Fig. 5(A). In Fig. 5(B), the percentage of phase-separated vesicle hydrated with 10mM NaCl solution is presented for fixed fraction of DOPC=40% and Chol=20%. As shown in Fig. 5(B), the phase separation changes with DPPG$^{(-)}$ concentration. Without salt, the phase boundary between L$_o$/L$_d$ two-phase coexistence, and L$_o$/L$_d$/S$_o$ three-phase coexistence, is positioned at DPPC/ DPPG$^{(-)}$= 25:15. On the other hand, in 10mM NaCl solution, the phase boundary is DPPC/ DPPG$^{(-)}$= 20:20. The phase boundary between the L$_o$/L$_d$/S$_o$ three-phase coexistence and L$_d$/S$_o$ or L$_o$/S$_o$ two-phase coexistence, also depends on the salt condition: the boundaries are DPPC/ DPPG$^{(-)}$= 20:20 (without salt) and 15:25 (10mM NaCl). These results suggest that the addition of salt affects phase structure of DOPC/DPPC/ DPPG$^{(-)}$/Chol mixtures.

## Discussion

One of our important results is that when neutral lipids are replaced by charged ones, the phase separation was suppressed for the DOPG$^{(-)}$/DPPC mixtures, whereas it was enhanced for mixtures of DOPC/ DPPG$^{(-)}$. Furthermore, by adding salt, these two mixtures approached the behaviour of the non-charged DOPC/DPPC mixture. As mentioned above, it was reported in the past experiments[19,21,22,23] that phase separation of other mixtures containing negatively charged unsaturated lipids was suppressed similarly to our DOPG$^{(-)}$/DPPC result. However, the enhanced phase separation for DOPC/ DPPG$^{(-)}$ is novel and unaccounted for.

We discuss now several theoretical ideas that are related to these empirical observations based on a phenomenological free energy model[19,20,27,28]. The first step is to take into account only the electrostatic contribution to the free energy, $f_{el}$, using the Poisson-Boltzmann (PB) theory. For symmetric monovalent salts (e.g., NaCl), the electric potential $\Psi(z)$ at distance $z$ from a charged membrane satisfies the PB equation:

$$\frac{d^2\Psi}{dz^2} = \frac{2en_b}{\varepsilon_w}\sinh\frac{e\Psi}{k_BT} \quad , \qquad \text{-(1)}$$

where $e$ is the electronic charge, $n_b$ the bulk salt concentration, and $\varepsilon_w$ the dielectric constant of the aqueous solution, $k_B$ the Boltzmann constant, and $T$ the temperature. For a charged membrane with area fraction $\phi$ of negatively charged lipids, the surface charge density is written as $\sigma = -e\phi/\Sigma$. The cross-sectional area $\Sigma$ of the two lipids is assumed, for simplicity, to be the same. The PB equation (1) can be solved analytically by imposing $\sigma$ as the electrostatic boundary condition, and the resulting electrostatic free energy is obtained as[29]

$$f_{\text{el}}(\phi) = \frac{2k_BT}{\Sigma}\phi\left[\frac{1-\sqrt{1+(p_0\phi)^2}}{p_0\phi} + \ln(p_0\phi + \sqrt{1+(p_0\phi)^2})\right], \quad -(2)$$

where $p_0 = 2\pi l_B l_D/\Sigma$ is a dimensionless parameter proportional to the Debye screening length $l_D = \sqrt{\varepsilon_w k_B T/2e^2 n_b}$, and to $1/\Sigma$, while $l_B = e^2/(4\pi\varepsilon_w k_B T) \approx 7\text{Å}$ is the Bjerrum length.

One essential outcome of the PB model is that for any $p_0$, the electrostatic free energy $f_{\text{el}}$ increases monotonically as a function of $\phi$, and a large fraction of negatively charged lipid will increase the free energy substantially. This implies that any charged domain formed due to lipid/lipid lateral phase separation would cost an electrostatic energy. Hence, within the PB approach, the phase separation in charged/neutral mixtures of lipids should be suppressed (rather than enhanced) as compared with neutral ones. Indeed, phase diagrams calculated by using a similar PB approach clearly showed the suppression of the phase separation[19,20,30,31]

The above argument does not explain all our experimental findings. Mixtures containing negatively charged saturated lipids are found to enhance the phase separation, and indicate that there should be an additional attractive mechanism between charged saturated lipids to overcome the electrostatic repulsion. Indeed, the demixing temperature in the DOPC/DPPG[(-)] mixture (Fig. 1) was found to be even higher than the chain melting temperature of pure DPPG[(-)] ($T_m$=41°C). Furthermore, the charged DPPG[(-)]/Chol binary mixtures exhibited the phase separation, whereas the neutral DPPC/Chol mixtures (see Fig. 2) did not.

The next step is to include entropic and enthalpic terms in the free energy for a membrane consisting of a mixture of negatively charged and neutral lipids,

$$f_{\text{tot}} = \frac{k_BT}{\Sigma}\left[\phi\ln\phi + (1-\phi)\ln(1-\phi) + \chi\phi(1-\phi)\right] + f_{\text{el}}, \quad -(3)$$

where the first and second terms in the square brackets account for the entropy and enthalpy of mixing between the charged and neutral lipids, respectively, while the last term, $f_{\text{el}}$, is the electrostatic free energy as in Eq. (2). As before, $\phi$ is the area fraction of the negatively charged lipid, $1-\phi$ is that of neutral lipid, and $\chi$ is a dimensionless interaction parameter between the two lipids (of non-electrostatic origin). Note that we took for simplicity the cross-sectional area $\Sigma$ of the two lipids to be the same, meaning that $\phi$ can be thought of as the charged lipid mole fraction. We note that the free energy formulation as in Eq. (3) was used in other studies, such as surfactant adsorption at fluid-fluid interface[32] or lamellar-lamellar phase transition[33]. In the case of a neutral lipid mixture membrane ($f_{\text{el}}=0$), this model leads to a lipid/lipid demixing curve with a critical point located at $\phi_c = 0.5$, $\chi_c = 2$.

The phase behaviour difference between mixtures of DOPC/DPPG[(-)] and DOPG[(-)]/DPPC also suggests a specific attractive interaction between DPPG[(-)] molecules. This is not accounted for by the PB theory of Eq. (2), but the enhanced phase separation can effectively be explained in terms of an increased $\chi$-value in Eq. (3) for mixtures containing DPPG[(-)]. We plan to explore the origins of such non-electrostatic attractive contributions in a future theoretical study, and in particular, to explore the relationship between the electrostatic surface pressure and the phase separation[34,35].

Although DOPG[(-)]/DPPC and DOPC/DPPG[(-)] mixtures look very similar from the electrostatic point of view, it is worthwhile to point out some additional difference between these mixtures (beside the value of the $\chi$ parameter). In particular, the phase behavior of DOPC/DPPG$^{(-)}$ approaches that of neutral DOPC/DPPC system by adding salt. Since the attractive force between DPPG$^{(-)}$ molecules vanishes by the addition of salt, we consider that this attractive force may be related to the charge effect. Because DOPG$^{(-)}$ has an unsaturated bulky hydrocarbon tail, its cross-sectional area $\Sigma$ is larger than that of DPPG$^{(-)}$ that has a saturated hydrocarbon tail. In the literature, the cross-sectional areas of DOPG$^{(-)}$ and DPPG$^{(-)}$ are reported to be 68.6Å$^2$ (at $T$=30°C) and 48Å$^2$ (at $T$=20°C), respectively[36]. This area difference affects the surface charge density $\sigma = -e\phi/\Sigma$. As a result, the counterion concentration near the charged membrane are different for DOPG$^{(-)}$/DPPC as compared with DOPC/DPPG$^{(-)}$. Based on the PB theory, Eq. (1), one can obtain the counterion concentration $n_0 = n^+(z \to 0)$, adjacent to the membrane

$$n_0 = n_b \left( p_0 \phi + \sqrt{(p_0 \phi)^2 + 1} \right)^2 \quad . \qquad \text{-(4)}$$

This relation is known as the Grahame equation[37,38], and is used in Fig. 6 to plot $n_0$ for $n_b$=10mM. As shown in Fig. 6(A), $n_0$ sharply increases when the cross-sectional area $\Sigma$ decreases. This tendency is significantly enhanced at higher area fraction $\phi$ of the charged lipid. In Fig. 6(B), $n_0$ is plotted for $\Sigma$ =50 Å$^2$ (solid line) and 70Å$^2$ (dashed line), which to a good approximation correspond to the values of DPPG$^{(-)}$ and DOPG$^{(-)}$, respectively. The larger value of $n_0$ for DPPG$^{(-)}$ may influence the relative domain stability that cannot be described by the simple continuum PB theory. We also speculate that the hydrogen bonds between charged head groups and water molecules can be affected by the presence of a large number of counterions. Although this counter-ion condensation is one of the possible explanations for the strong attraction between DPPG$^{(-)}$ molecules, it is not enough in order to describe the underlying mechanism completely. In addition, it is important to understand whether this attractive force is also observed in systems including other types of charged lipids (e.g. phosphatidylserine (PS$^{(-)}$)). Such questions remain for future explorations.

Moreover, we found that ternary mixtures of DPPC/DPPG$^{(-)}$/Chol exhibit phase separation between DPPC/Chol-rich and DPPG$^{(-)}$-rich phases. This is because the strong attraction between DPPG$^{(-)}$ molecules excludes cholesterol from DPPG$^{(-)}$-rich domains. In addition, the difference of the molecular tilt between different lipids may also affect this phase separation. The localization of cholesterol strongly depends on the molecular shape of membrane phospholipids. It was reported that polar lipids, such as DPPC, which contain both positively and negatively charges in their head group, tend to tilt due to electrostatic interaction between the neighboring polar lipids[39,40]. The tilting produces an intermolecular space that cholesterol can occupy. However, since the molecular orientation of DPPG$^{(-)}$ is almost perpendicular to the membrane surface, it will be unfavorable for cholesterol to occupy such a narrow space between neighboring DPPG$^{(-)}$ molecules.

The three phase coexistence in four-component mixtures of DOPC/DPPC/DPPG$^{(-)}$/Chol=40:20:20:20 could be caused by the same mechanism. Unsaturated DOPC forms an L$_d$ phase, whereas cholesterol, which is localized in DPPC domains, forms an L$_o$ phase. Thus, the DPPG$^{(-)}$-rich region results in an S$_o$ phase. Since the hydrocarbon tails of DPPG$^{(-)}$ in the S$_o$ phase are highly ordered, whereas the DOPC hydrocarbon tails in the L$_d$ phase are disordered, the S$_o$/L$_d$ line tension is larger than the line tension of the S$_o$/L$_o$ interface. Therefore, S$_o$ domains are surrounded by L$_o$ domains in order to prevent a direct contact between S$_o$ and L$_d$ domains.

Although charged lipids in biomembranes are generally assumed to be in the fluid phase, the S$_o$ phase with a large amount of charged lipids is observed in our experiments (on 4-component mixtures). Notably, the formation of the S$_o$ phase has been reported in model membrane systems either by decreasing the cholesterol fraction or by increasing the membrane surface tension[7,8]. Although the $S_o$ phase has not been seen *in vivo*, we believe that our study on model membrane is meaningful and will help to reveal some important physicochemical mechanisms that underlie the phase behaviour and domain formation of lipid membranes *in vivo*. The $L_o$ domains in artificial membranes can be regarded as models mimicking rafts in biomembranes. Because most of proteins have electric charges, sections of the proteins that have positive charges can easily be attached to the negatively charged domains due to electrostatic interactions. Conversely, negatively charged sections of proteins are electrically excluded from such domains. Thus, such charged domains may play an important role in the selective adsorption of charged biomolecules.

Finally, we comment that, in all of our experiments, the salt concentration was 10mM. This concentration is lower than the concentration in physiological conditions of living cells, where the monovalent salt concentration is about ~140mM. From our results, we can see that screening by the salt is significant even for 10mM[19,20,30,31].

## Conclusions

In the present study, we investigated the phase separation induced by negatively charged lipids. As compared to the phase-coexistence region (in the phase diagram) of neutral DOPC/DPPC mixtures, the phase separation in the charged DOPG$^{(-)}$/DPPC case is suppressed, whereas it is enhanced for the charged DOPC/DPPG$^{(-)}$ system. The phase behaviours of both charged mixtures approach that of the neutral mixture when salt is added due to screening of electrostatic interactions. In DPPC/DPPG$^{(-)}$/Chol ternary mixtures, the phase separation occurs when the fraction of charged DPPG$^{(-)}$ is increased. This result implies that cholesterol localization is influenced by the head group structure as well as the hydrocarbon tail structure. Furthermore, we observed three-phase coexistence in four-component DOPC/DPPC/DPPG$^{(-)}$/Chol mixtures, and that the phase-separation strongly depends on the amount of charged DPPG$^{(-)}$.

Our findings shed some light on how biomembranes change their own structures, and may help to understand the mechanisms that play an essential role in the interactions of proteins with lipid mixtures during signal transduction.

## Acknowledgements

Technical assistance from Ms. Ryoko Sugimoto and Mr. Masato Amino is greatly appreciated. We thank Mr M. C. Blosser, Dr R. Dimova, Dr M. Hishida, Dr W. Shinoda and Dr T. Taniguchi for fruitful discussions and comments. This work was supported in part by the Sasagawa Scientific Research Grant from The Japan Science Society and Young Scientist (B) from JSPS and on Priority Areas "Molecular Robotics" and "Spying Minority in Biological Phenomena" from MEXT and Kurata Grant from the Kurata Memorial Hitachi Science and Technology Foundation. SK acknowledges support from the Grant-in-Aid for Scientific Research on Innovative Areas "Fluctuation & Structure" (grant No. 25103010), Grant-in-Aid for Scientific Research (C) grant No. 24540439 from the MEXT of Japan, and the JSPS Core-to-Core Program "International research network for non-equilibrium dynamics of soft matter". DA acknowledges support from the Israel Science Foundation (ISF) under grant no. 438/12.

## Notes and references

[a] School of Materials Science, Japan Advanced Institute of Science and Technology, 1-1 Asahidai, Nomi city, Ishikawa, 923-1292, Japan. E-mail: t-hamada@jaist.ac.jp; Tel: +81 761-51-1670

[b] Department of Chemistry, Graduate School of Science and Engineering, Tokyo Metropolitan University, Tokyo 192-0397, Japan

*c Raymond and Beverly Sackler School of Physics and Astronomy, Tel Aviv University, Ramat Aviv 69978, Tel Aviv, Israel*
† *These authors contributed equally to this work.*

**Table.1** The four neutral and negatively charged lipids and their chain melting temperatures

|  | Neutral head (PC) | Negative charged head (PG) |
|---|---|---|
| Saturated tail (DP) | DPPC $T_m$= 41℃ | DPPG$^{(-)}$ $T_m$= 41℃ |
| Unsaturated tail (DO) | DOPC $T_m$= -20℃ | DOPG$^{(-)}$ $T_m$= -18℃ |

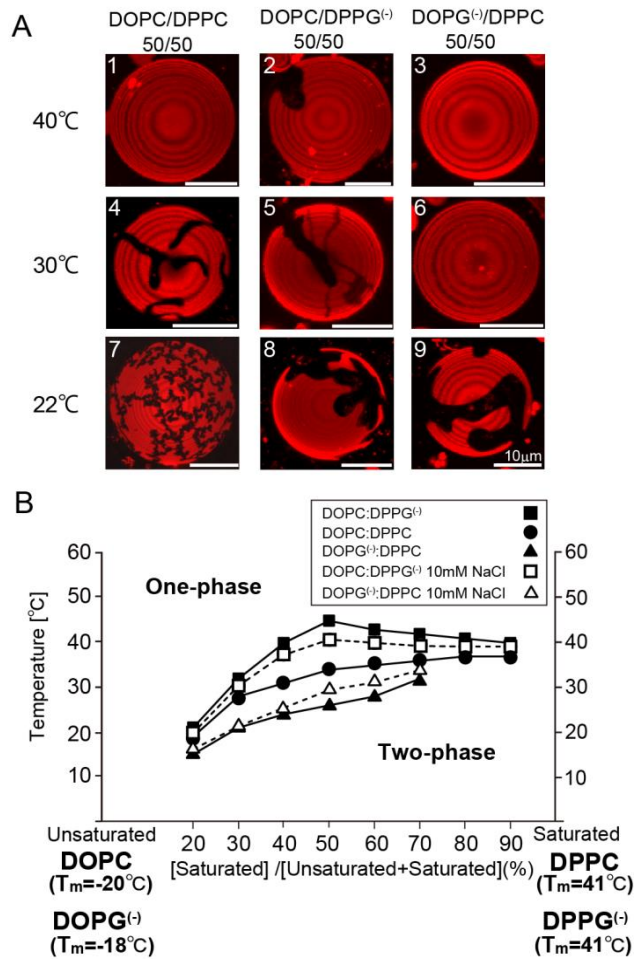

**Fig.1** Phase behaviour in binary lipid mixtures (DOPC/DPPC, DOPC/DPPG(-), DOPG(-)/DPPC). (A) Microscopic images of the phase separation for three temperatures, 22℃, 30℃ and 40℃. Red and black regions indicate unsaturated lipid-rich ($L_d$) and saturated lipid-rich ($S_o$) phases, respectively. (B) Phase boundary (miscibility temperature) between one-phase and two-phase regions (filled square: DOPC/DPPG(-), filled circle: DOPC/DPPC, filled triangle: DOPG(-)/DPPC, open square: DOPC/DPPG(-) in 10mM NaCl, open triangle: DOPG(-)/DPPC in 10mM NaCl).

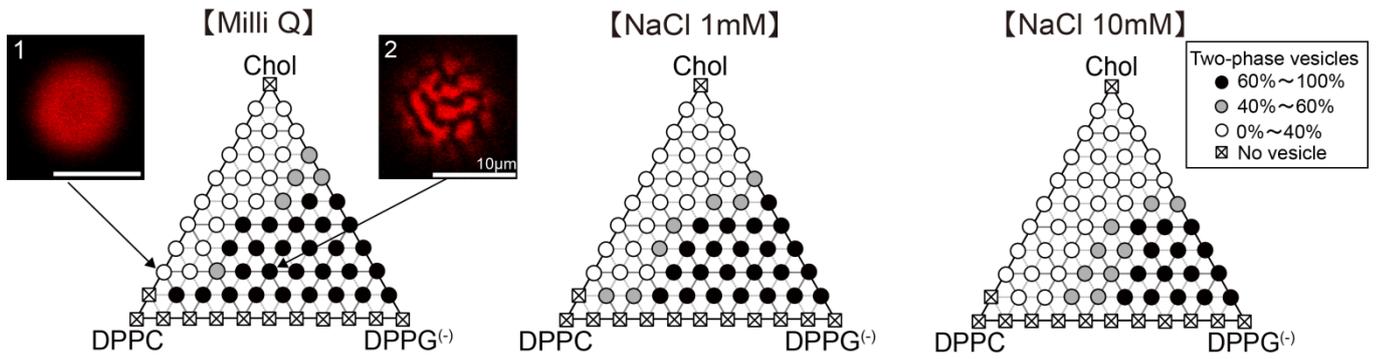

**Fig.2** Phase diagrams of DPPC/DPPG(-)/Chol mixtures in Milli Q and NaCl solutions (left: Milli Q, centre: NaCl 1mM, right: NaCl 10mM) at room temperature (~22°C). Filled, grey, and open circles correspond to systems where 60-100%, 40-60%, and 0-40% of the vesicles, respectively, exhibit two-phase regions. Microscopic images of GUVs are taken at composition of DPPC/Chol=80/20 (image 1) and DPPC/DPPG(-)/Chol=40/40/20 (image 2) in Milli Q water at 22°C. Cross marks indicate the region where the vesicles formed by natural swelling method are not stable.

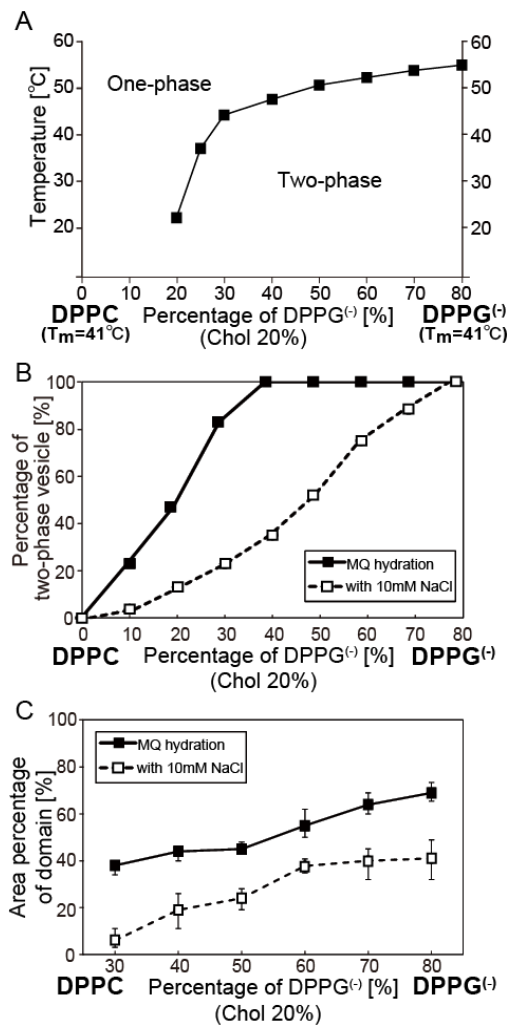

**Fig.3** (A) Phase diagram of DPPC/DPPG$^{(-)}$/Chol mixtures for fixed Chol = 20%. (B) Percentage of two-phase vesicle at 22 °C, and (C) area percentage of the S$_o$ phase at 22 °C as a function of DPPG$^{(-)}$/DPPC ratio for fixed Chol = 20%. Filled and open squares indicate Milli Q and 10mM NaCl solution, respectively.

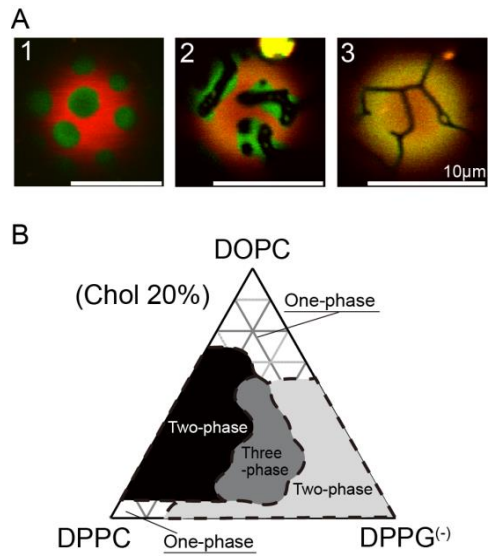

**Fig.4** (A) Phase behaviour in multi component mixtures of DOPC/DPPC/DPPG$^{(-)}$/Chol. (A) Microscope images of GUVs at compositions of DOPC/DPPC/Chol = 40/40/20 (image 1), DOPC/DPPC/DPPG$^{(-)}$/Chol = 40/20/20/20 (image 2), and DOPC/DPPG$^{(-)}$/Chol=40/40/20 (image 3) at 22 ℃. Red, green, and dark regions indicate DOPC-rich ($L_d$), DPPC/Chol-rich ($L_o$), and DPPG$^{(-)}$-rich ($S_o$) phases, respectively. The yellow region in image 3, which includes a large amount of DOPC and Chol indicates an $L_d$ phase. (B) Phase diagram of four-component mixtures of DOPC/DPPC/DPPG$^{(-)}$/Chol for fixed Chol=20% at 22 ℃. Black, grey, and light grey regions denote, respectively, $L_o/L_d$ two-phase coexistence, $L_o/L_d/S_o$ three-phase coexistence, and $L_d/S_o$ or $L_o/S_o$ two-phase coexistence.

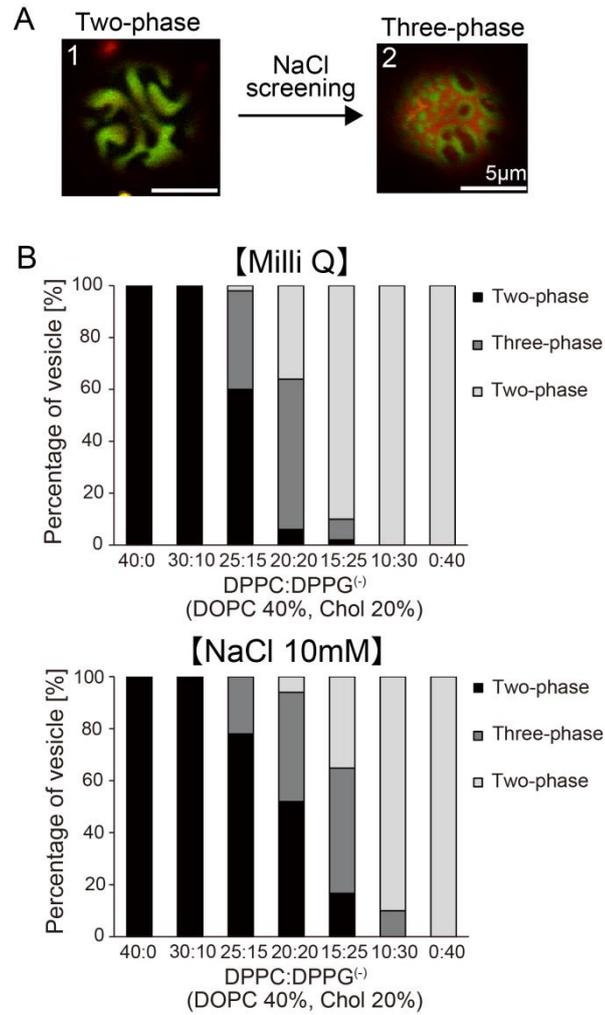

Fig.5 (A) Fluorescence microscopy images of phase separation in DOPC/DPPC/DPPG(-)/Chol=40:15:25:20 hydrated by Milli Q water (image 1) and 10mM NaCl solution (image 2) at 22 ℃. (B) The phase diagram of four-component mixtures hydrated by Milli Q water (upper graph) and 10mM NaCl solution (lower graph), respectively. Temperature was fixed at 22 ℃. The relative ratio between DPPG(-) and DPPC is changed while keeping fixed amount of DOPC=40% and Chol=20%. Black, grey, and light grey regions indicate the $L_o/L_d$ two-phase coexistence, $L_o/L_d/S_o$ three-phase coexistence, and $L_d/S_o$ or $L_o/S_o$ two-phase coexistence, respectively.

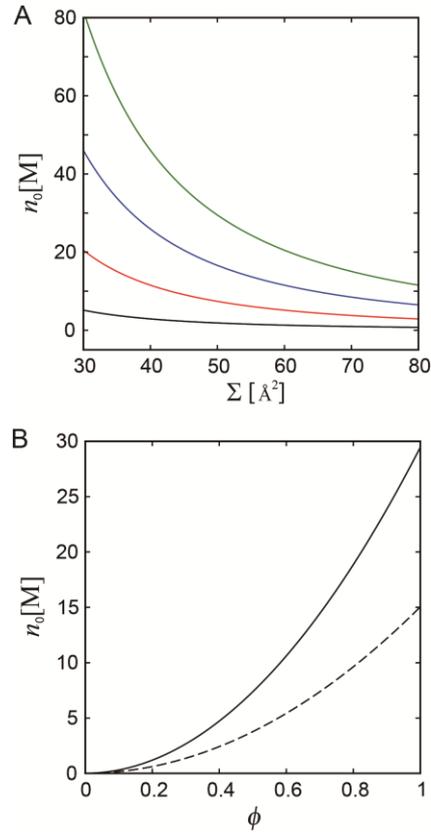

**Fig.6** (A) The counterion concentration, $n_0 = n^+(z \to 0)$, extrapolated to the membrane vicinity as a function of cross-sectional area per lipid $\phi$ for the bulk salt concentration, $n_b = 10\text{mM}$. The different line colours represent $\phi = 0.25$ (black), $0.5$ (red), $0.75$ (blue), and $1.0$ (green). (B) The counterion concentration at the membrane as a function of the charged lipid concentration, for bulk salt concentration, $n_b = 10\text{mM}$. The solid and dashed lines denote $\Sigma = 50$ Å² and $70$ Å², respectively.